\documentclass[10pt, twocolumn]{IEEEtran}
 \usepackage{amsmath,amssymb}
 \usepackage{subfigure}
 \usepackage{graphicx,graphics,color,psfrag}
 \usepackage{cite,balance}
 \allowdisplaybreaks
 \usepackage{algorithm}
 \usepackage{accents}
 \usepackage{amsthm}
 \usepackage{bm}
 \usepackage{algorithmic}
 \usepackage[english]{babel}
 \usepackage{multirow}
 \usepackage{enumerate}
 \usepackage{cases}
 \usepackage{stfloats}
 \usepackage{dsfont}
 \usepackage{color,soul}
 \usepackage{amsfonts}
 \usepackage{cite,graphicx,amsmath,amssymb}
 \usepackage{subfigure}
 \usepackage{fancyhdr}
 \usepackage{hhline}
 \usepackage{graphicx,graphics}
 \usepackage{array,color}

 \usepackage{amsmath}

\usepackage{lipsum}
\usepackage{mwe}

\usepackage[utf8]{inputenc}
\usepackage{graphicx}
\usepackage{cleveref}

\usepackage{multirow,makecell}
\usepackage{graphicx}
\usepackage{tabularx, booktabs}


\newcounter{parentnumber}

\def\phi{\varphi}

\def\({\left(}
\def\){\right)}

\def\b0{{\mathbf{0}}}

\begin{document}

\title{\Huge Understanding Uncertainty of Edge Computing: New Principle and Design Approach}

\author{Sejin Seo, Sang Won Choi, Sujin Kook, Seong-Lyun Kim, and Seung-Woo Ko   \\
\thanks{Sejin Seo, Sujin Kook, and Seong-Lyun Kim are with the School of Electrical and Electronic Engineering, Yonsei University, Seoul, Korea (email: {sjseo, sjkook, slkim}@ramo.yonsei.ac.kr); Sang Won Choi is with Korea Railroad Research Institute, Uiwang, Korea (email: swchoi@krri.re.kr); and Seung-Woo Ko is with  the Division of Electronics and Electrical Information Engineering at Korea Maritime and Ocean University, Busan, Korea (email: swko@kmou.ac.kr).}
}
\maketitle

\begin{abstract} 
Due to the edge's position between the cloud and the users, and the recent surge of deep neural network (DNN) applications, edge computing brings about uncertainties that must be understood separately. Particularly, the edge users' locally specific requirements that change depending on time and location cause a phenomenon called \emph{dataset shift}, defined as the difference between the training and test datasets' representations. It renders many of the state-of-the-art approaches for resolving uncertainty insufficient. Instead of finding ways around it, we exploit such phenomenon by utilizing a new principle: \emph{AI model diversity}, which is achieved when the user is allowed to opportunistically choose from multiple AI models. To utilize AI model diversity, we propose \emph{Model Diversity Network} (MoDNet), and provide design guidelines and future directions for efficient learning driven communication schemes. 
\end{abstract}

\section{Introduction}\label{Introduction}
The strive for making lighter consumer devices equipped with the access to unlimited applications pushed the resources to servers in the cloud. Now, the lack of bandwidth and the demand for real-time and user specific services are pulling the resources back to the network \emph{edge}. The edge, which primarily served to act as access points for the radio, tries to expand its role as storage and computing servers for user specific needs \cite{b1}. With this evolution, more diverse computation tasks are demanded from the edge, ranging from traditional tasks with deterministic results to tasks that utilize the recent advances in \emph{machine learning} (ML) and related developments of \emph{artificial intelligence} (AI) solutions \cite{b2}. The reason behind this is that AI solutions, especially \emph{deep neural networks} (DNN) models, strongly embody the specific features required by a task. Complying with the demand for more complex tasks notwithstanding the edge’s rather limited computation capacity, DNN based AI solutions that used to belong to the realm of heavy computing servers are now transformed to more compact versions and implemented in {mobile edge computers} and even in light consumer devices \cite{b3}. 

Despite these efforts, the required performance may not be guaranteed, because all models are imperfect embodiment of the reality. This is intensified for the compact models that are used at the edge, because they lack the ability to represent the complex features of the reality. Moreover, the data at the edge are noisy, sparse, and biased, causing the degradation of performance while also increasing the variance of the result. The variance of performance is understood as the \emph{uncertainty} of the performance, which becomes fatal for tasks that are life-and-death related. For example, this is a serious issue for autonomous driving: {Tesla and Uber autopilot accidents all resulted due to poor reflection of the visual factors that should have been considered, e.g. concrete barriers and data exposed to sunlight.}

In this article, we define computation uncertainty from the perspective of edge computing. To the best of our knowledge, it is the first work to address this issue comprehensively. Specifically, we summarize the common factors and conditions that cause computation uncertainty, including uncertainties in cloud servers, DNN models, and the edge. In the computer science literature, most works focus on designing ways to represent all relevant features into a single AI model by increasing its complexity, which is infeasible at the edge. Instead, we aim at choosing the best option available among multiple lighter AI models, defined as AI model diversity. To this end, we propose a novel edge computing architecture, which realizes AI model diversity in an edge environment relying on wireless access. Lastly, we end by providing several interesting directions for optimizing the architecture from the wireless communication perspective. 

\section{Computation Uncertainty of Edge Computing}\label{Sec: CompUnc}

\subsection{Computation Uncertainty}
Computation uncertainty is the variance of a computation result. The uncertainty is caused by various reasons, and they are classified differently according to the application of concern. In cloud computing, they are classified into two groups: parametric uncertainties and system uncertainties \cite{b4}. Parametric uncertainties are caused by environmental parameters that cannot be reduced directly by adjusting the system. {On the other hand, system uncertainties are related to factors that can be controlled by the system, e.g. by acquiring additional information or designing a more adequate system.} In \cite{b4}, different causes of uncertainties are listed and some methods for resource provisioning and scheduling under uncertainty are provided.  

As DNNs are gaining unprecedented amount of attention, computation uncertainty is being defined specifically for them. In the ML literature, the uncertainties are classified as either aleatoric or epistemic. First, aleatoric uncertainty is the uncertainty in the data caused by inherent randomness in the generation of the data. Second, epistemic uncertainty is the uncertainty in the model caused by insufficient training data for representing all necessary features of the input domain. The uncertainties are often linked with the level of trust we can put into a model. This notion is drawing more attention as DNN is expanding its applications to those that require high level of reliability, thus requiring not only the point accuracy of a model but the confidence in a model \cite{b5}.

It is worth noting the relationship between the training data and model uncertainty. When the data are sparse, the model uncertainty is significantly increased, because ML models are heavily dependent on the training data provided to them.  For DNN, this tendency worsens, because DNNs are excellent interpolators but poor extrapolators of knowledge.

\begin{figure}[t]
\centerline{\includegraphics[width=8cm]{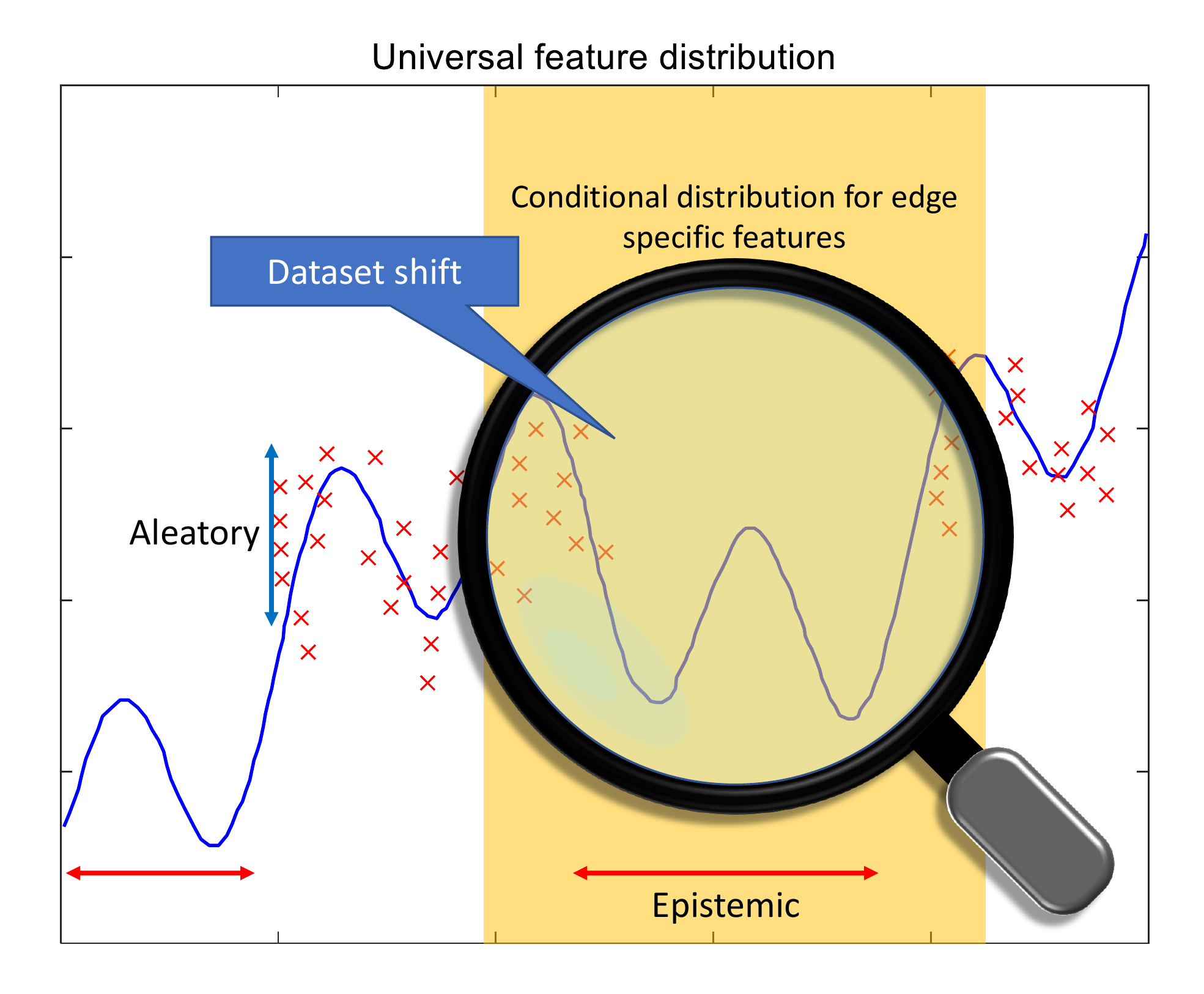}}
\caption{One-dimensional illustration of computation uncertainties and dataset shift at the edge.}
\label{fig1}
\end{figure}

\subsection{Challenges at the Edge}

Uncertainty has to be understood separately for the edge, because it inherits uncertainties from other areas and faces new challenging uncertainties as well. First of all, the edge inherits the uncertainties present in cloud computing, because edge computing is similar in structure to cloud computing, except it has more stringent requirements and limitations. In addition, the edge inherits the uncertainties present in DNN as well, when the edge utilizes DNN models. We follow the conventions of the ML literature, which encapsulates the uncertainties at the edge well.

Aside from these shared difficulties, the edge is posed with a more challenging situation due to the locally specific requirements like user, location, and time specific requirements. The bias incurred by the locally specific requirements makes the target distribution deviate from the universal distribution, i.e. distribution representing the entire feature space. This deviation is precisely described with the phenomenon termed \emph{dataset shift}, which happens when the joint distribution of the input  and output data are different for the training and test stages \cite{b6}. As shown in Fig. \ref{fig1}, the prevalence of dataset shift in edge computing greatly increases the level of uncertainty. 

Let us revisit the autopilot problem with this perspective. Consider that a universal DNN model is used for the autopilot. This model is trained for diverse situations, giving it very high accuracy on average. However, if this model is not trained comprehensively for situations with high level of sunlight, the model’s performance becomes extremely uncertain for situations where the camera sensor only takes in images exposed to intense sunlight. Now, if the designers become aware of this gap, they can solve this local problem by training the model further with these new images exposed to intense sunlight. Nonetheless, if these adjustments are not made in real-time, then accidents become inevitable. To make things worse, although heavy DNN models are required to represent all these diverse features simultaneously, they cannot be run at the edge due to computation capacity limitation. As a result, compact models have to be used, but they further increase the uncertainty due to their inherent lack of capabilities to represent complex features.

\subsection{The State-of-the-Art}
In computer science, numerous efforts have been made to overcome performance degradation due to computation uncertainty. Firstly, most works that address data uncertainty try to understand the noise factors as extra features for their models. These works either try to filter the noise to recover the original data or include it into their models to avoid performance degradation. For example, deep ensembles use distributional parameter estimation to understand the variance of the output \cite{b7}, and \emph{denoising autoencoder} (DAE) can be trained to recover the uncorrupted data by intentionally training them with noise \cite{b8}.

Secondly, to address the model uncertainty, most approaches utilize randomness in their models and exploit the ensemble average. These randomization methods either work to add complexity into their models to address underfitting issues or regularization to avoid overfitting. An approach called \emph{Bayesian neural networks} (BNN) use prior distributions to generate the weight parameters randomly instead of using deterministic weights to represent uncertainty \cite{b5}, but they require heavy computation and do not scale well. Another approach that tries to overcome such shortcomings is dropout, which randomly omits weight parameters to perform variational Bayesian approximation \cite{b9}. Though useful, they are not feasible at the edge because they either fall short of decreasing the complexity of the models or suffer performance degradation when faced with significant dataset shift, all because they still rely on the expectation of the dataset.

As shown in Table \ref{my-label}, the approaches that deal with data uncertainty, i.e. deep ensembles and DAE, do not deal with model uncertainty; the approaches that deal with model uncertainty, i.e. BNN and dropout, do not deal with data uncertainty; also, all approaches except dropout are not scalable for time, location, and user dynamics, because their model complexity increases along with those factors. To this end, new ways to overcome uncertainty at the edge must be devised.

\begin{table*}
 \caption{The state-of-the-art approaches: addressing problems at the edge}
\label{my-label}
\begin{tabularx}{\textwidth}{@{}l*{10}{X}c@{}}
\toprule
Approach     & Data Uncertainty & Model Uncertainty & Scalability & Dynamic Dataset Shift \\ 
\midrule
Deep ensembles \cite{b7}   & O & X & X & Partially \\ 
DAE \cite{b8}              & O & X & X & Partially \\ 
BNN \cite{b5}              & X & O & X & X \\ 
Dropout \cite{b9}          & X & O & O & Partially   \\ 
\addlinespace
\textbf{MoDNet (proposed)} & \textbf{O} & \textbf{O} & \textbf{O} & \textbf{Exploited}  \\ 
\bottomrule
\end{tabularx}
\end{table*}

\section{AI Model Diversity: Exploiting Dataset Shift}\label{Sec: Modnet}
From the data science perspective, the edge is characterized by the frequent occurrence of dynamic dataset shift, where the test data domain is often shifted depending on the environment. This occurs because the test data required by an edge user are biased or even bound to certain edge specific features, which may even be considered peripheral to a universal representation. Such deviance renders other state-of-the-art approaches insufficient for the edge, because they are all focused on building a more comprehensive representation of the entire test data of interest, regardless of the dataset shift. To resolve this issue, we introduce a new principle called \emph{AI model diversity}.

\subsection{AI Model Diversity}
Instead of passively resolving dataset shift, it is exploited by using AI model diversity granted by allowing the user to choose from multiple compact AI models that represent different edge specific features. Despite their limitations, even the compact models are capable when they only need to focus on the specific features that are magnified by the dataset shift. This simple but powerful scheme helps resolve data and model uncertainties, while ensuring scalability. 

\begin{itemize}
\item \textbf{Data uncertainty}: AI model diversity resolves data uncertainty by making each model train for the noise factors that are correlated with the edge. For instance, recurring background of an edge should be considered as a significant feature to be trained along with the object itself.
\item \textbf{Model uncertainty}: Using multiple AI models grants wider coverage of the feature space that should be represented for the edge. Also, it exploits the dataset shift to gain efficiency and simplicity. Whereas most neural networks are designed to represent a universal need, the compact models in an edge server efficiently covers the features dominant at the edge. This difference must be noted due to the edge’s constraints. 
\item \textbf{Scalability}: The proposed approach is very scalable for time, location, and user dynamics. Whereas adaptation is very costly for complex and heavy AI models, it can readily adapt to a changing environment by choosing or training a compact model that better represents the current situation. 
\end{itemize}

Recently, a few works explore diversity for ML regarding the data, model, and inference, e.g. \cite{b10}. However, similar to the state-of-the-art approaches, they do not address the complications and implementation issues related to the edge, which are this article’s major interest.

\begin{figure*}
\centering
    \includegraphics[width=17cm]{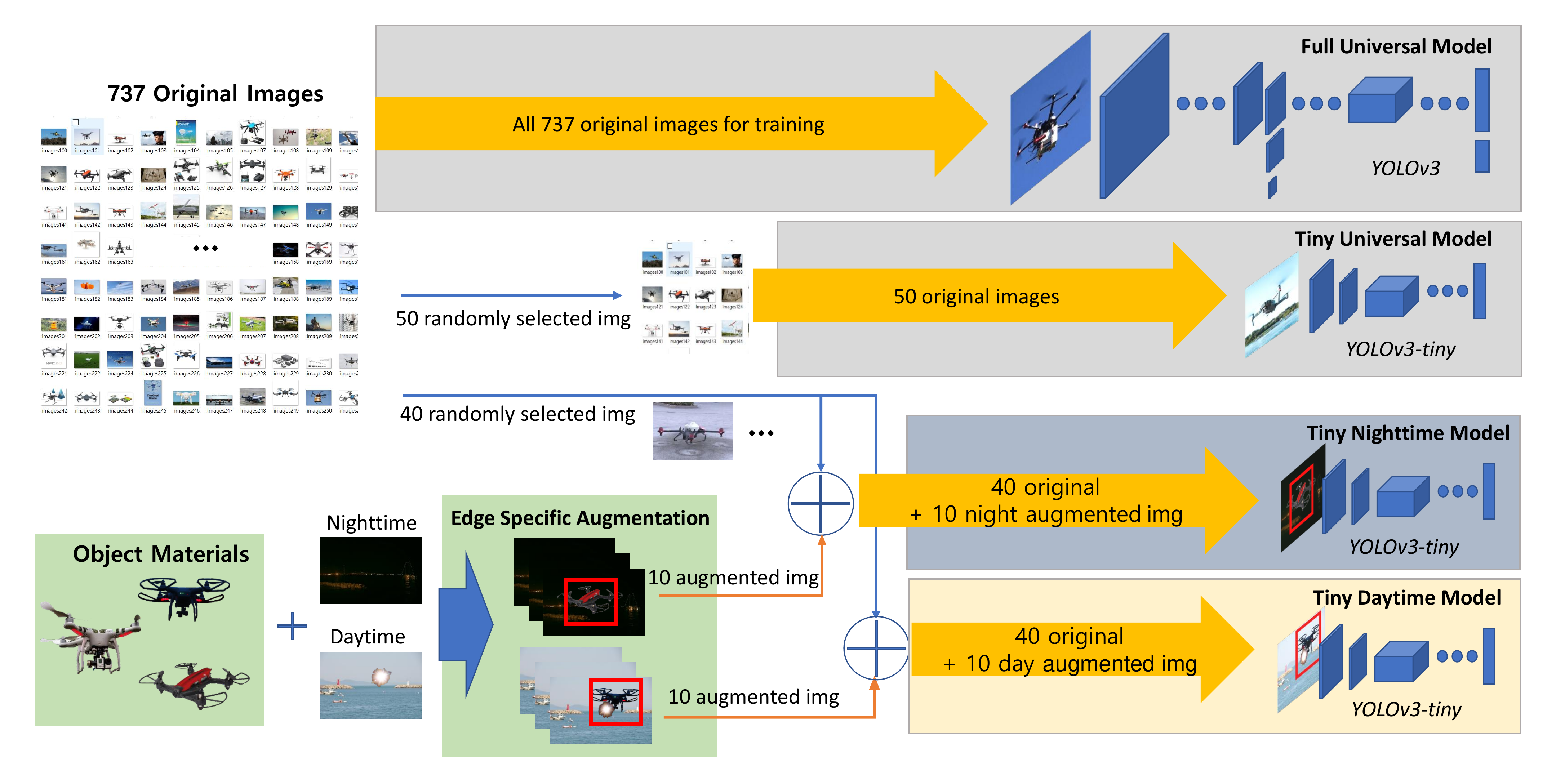}
    \caption{Graphical explanation of four AI model training used to verify AI model diversity. Further explanation and data are available at: https://github.com/kgbssj.} 
\label{fig2}
\end{figure*}

\subsection{Experimental Verification: Object Detection}\label{subsection:Experiment}

The following object detection experiment is conducted to illustrate and verify how the dynamic dataset shift at the edge can be exploited by using AI model diversity. The experiment uses state-of-the-art real-time object detection algorithms, i.e. \emph{You-Only-Look-Once} (YOLO) and YOLO-tiny models, to represent a universal and local model, respectively \cite{b11}. To explain in more detail, three tiny models that can be run on realistic edge computing devices (e.g. NVidia Jetson Xavier) are trained with scarce \emph{unmanned aerial vehicles} (UAV) images, while a full model that has to be trained in a heavier computing server (RTX 2080 Ti), i.e. YOLO, is trained with abundant UAV images.  

In the experiment, $737$ annotated original images containing one or more UAVs are used as training data. Additionally, edge specific models use manually augmented images along with the original images for training, as illustrated in Fig. \ref{fig2}. Lastly, the test dataset is exclusive from the training dataset. 

The four AI models are juxtaposed for comparison: a full universal model, a tiny universal model, a tiny nighttime specific model, and a tiny daytime specific model. The full universal model, which is likely to be run at the cloud, uses all $737$ training data. With RTX 2080 Ti, it takes approximately $3$ hours $45$ minutes to train. Considering the constraints at the edge, a tiny universal model is trained by randomly selecting $50$ images from the entire training dataset. The tiny model needs approximately $55$ minutes to train.  

The edge specific models are trained with $40$ images that are randomly selected from the original train dataset and $10$ images that are augmented to represent certain edge specific features. The tiny nighttime specific model is tuned by using $10$ UAV images (without background) augmented with a nighttime image. Similarly, the tiny daytime specific model is tuned by training $10$ augmented UAV images with sun flares to represent UAVs exposed to sunlight. These models are all trained for $6000$ iterations or less depending on the level of overfitting. Similar to the tiny universal model, both of the edge specific models need approximately $55$ minutes to train. 

\begin{figure}[t]
\centerline{\includegraphics[width=9.0cm]{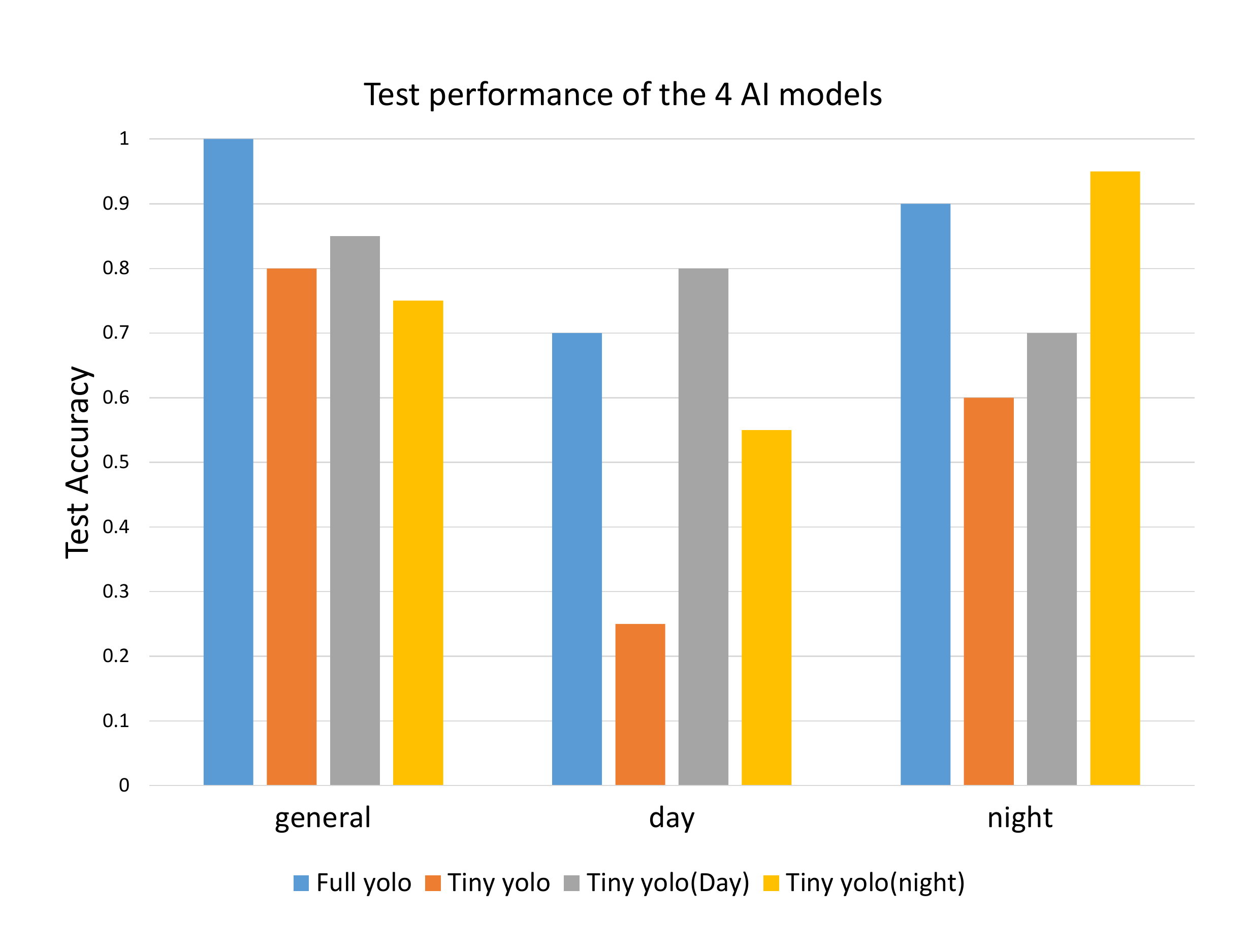}}
\caption{The average object detection accuracy of four AI models described in Fig. \ref{fig2} and  Sec. \ref{subsection:Experiment}.  }
\label{fig3}
\end{figure}

To evaluate the performance of each model, three test datasets with $20$ images each are used. The test datasets represent a general environment, nighttime environment, and daytime environment. From the graph in Fig. \ref{fig3}, it is evident that the universal model is nearly perfect for the general environment by using more depth and nodes, whereas all tiny models suffer from minor performance degradation due to the lack thereof. 


However, the full universal model’s performance plummets ($70$\%) for images exposed to sunlight during daytime. This is even worse for the tiny universal model ($25$\%), but it is very efficiently recovered for the tiny daytime specific model ($80$\%). This tendency continues for the tests during the nighttime; the tiny nighttime specific model is shown to outperform ($95$\%) the full universal model ($90$\%). These results demonstrate how the dynamic dataset shift directly impacts the performance. To explain, the dataset shift makes some features highly relevant at the edge, rather exclusively. These features should be exploited and less relevant features should be pruned to achieve higher accuracy without the expense of using larger weights or designing a specific architecture for each edge.  


\subsection{Realizing AI Model Diversity at the Edge: Data Augmentation and Fit Probability }

Due to data scarcity and dynamic dataset shift at the edge, two major steps are necessary for exploiting AI model diversity. The first step is to use appropriate data augmentation to train for the edge specific features to avoid overfitting to the training data. Sometimes, simple augmentation tools like flipping or darkening will suffice as in the object detection experiment. 
And yet, more sophisticated augmentation like random feature insertion or \emph{self-adversarial training} (SAT) may be necessary when a better performance is required. 

Now that the models are trained to satisfy some of the edge specific features, the second step is to take advantage of them. To exploit AI model diversity, it is crucial to obtain an estimation of the models’ accuracy. For this purpose, the \emph{fit probability} between the data and an AI model is defined as the probability that the program returns a correct answer. More precisely, the fit probability measures the distance between the feature distributions of the user’s test dataset and the model’s representation. If the latent distributions of the user’s needs and the model are known in advance, this could be understood as the \emph{Kullback-Leibler} (KL) divergence of these distributions. However, in a real system, these distributions are unknown, so the distance between them is estimated by sampling the model’s response to the train dataset, which acts as the surrogate for the actual feature distribution. Thus, the users augment annotated data based on the data augmentation process above and use them as pilot data to test and estimate the fit probability. It is analogous to transmitting pilot sequences for channel estimation in wireless communications, which aims at ensuring a desired data rate by evaluating the channel’s state. 

\begin{figure*}
\centering
    \includegraphics[width=17cm]{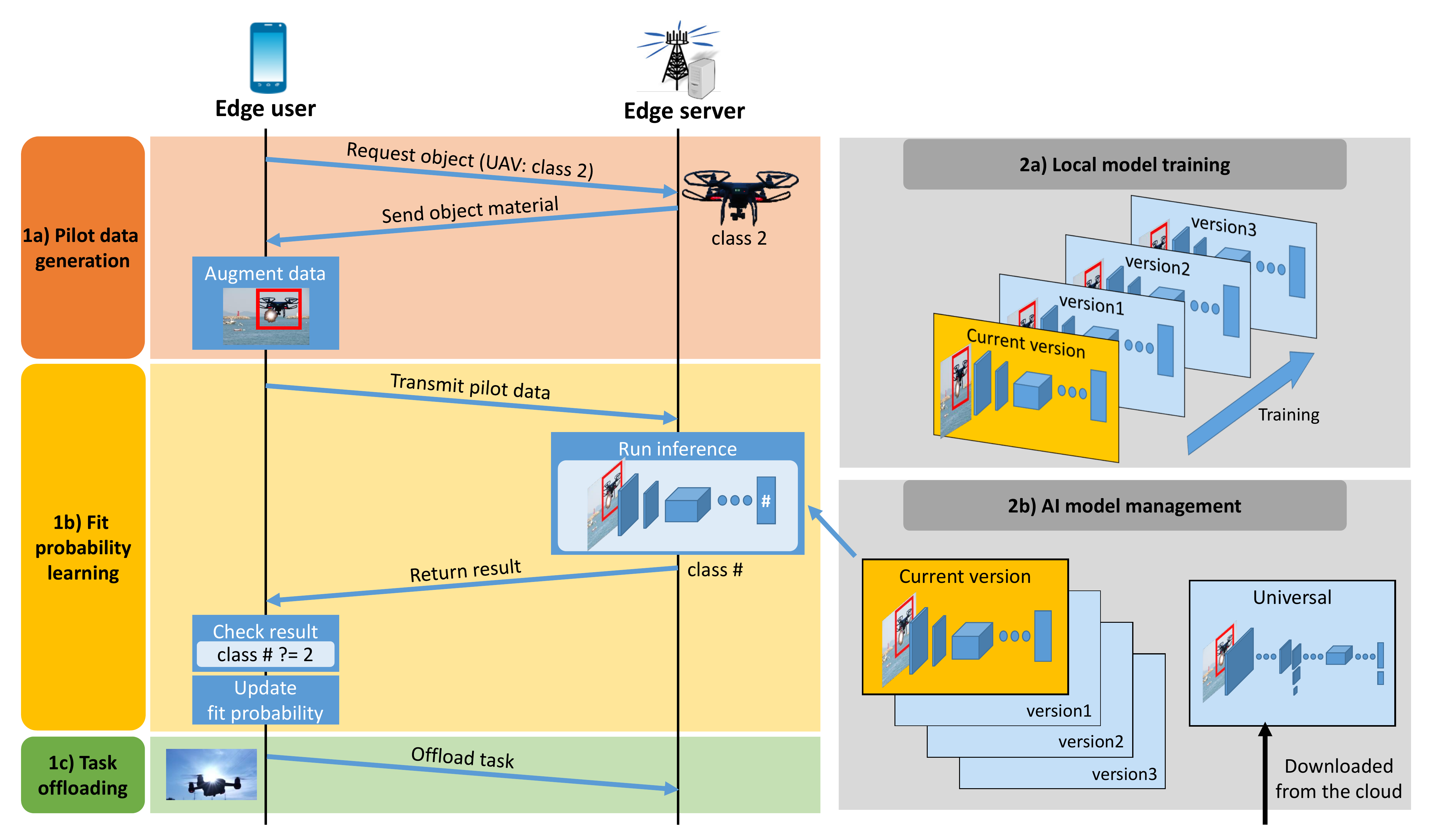}
    \caption{Task offloading and model training procedure of MoDNet.} 
\label{fig4}
\end{figure*}

\subsection{Edge Computing Network Architecture: MoDNet}\label{Subsection:Architecture}

To realize AI model diversity at the edge, we propose a new edge computing architecture called \emph{Model Diversity Network} (MoDNet), comprising edge users that act as the clients of computation tasks and edge servers that are linked with the edge users through wireless links; the edge servers have diverse AI models in them. Here, the tasks are parallel in meaning to the test data from above, but distinguished according to the edge computing perspective for more clarity. Due to their lack of computation capability, the edge users need to offload the tasks to the edge servers. There are different design issues in both sides of  edge users and servers, because users do not know how well an AI model would respond to their data and the servers do not know what kind of results the user would demand. To address the issues in both directions, the architecture is designed and described as follows: 

\subsubsection{\textbf{Task Offloading Process}}

To start, there are pre-trained AI models in the servers. To decide whether it is beneficial to offload the task, the edge user has to learn the fit probability between its task and the server’s AI models. To do so, these processes illustrated in Fig. 4 are necessary; and they all happen within a second.  

\begin{enumerate} [a)]
\item \emph{Pilot data generation}: When the user requests for a class of object, the server returns the user with generic object material that does not contain any edge specific features. The user augments the generic material with edge specific features, e.g. background and lighting. These pilot data do not need manual annotation, because the material intrinsically contains its annotation. 

\item \emph{Fit probability learning}: The user transmits the pilot data to determine whether the AI model in the edge server is fit for its tasks. Using the pilot data, the server samples the model's response and compares the result with its annotation. It is called ``success" if the result is correct. The comparison result is returned back to the user.  The user then estimates the fit probability by computing the ratio of the number of success events to the total number of transmitted pilot data. 

\item \emph{Task offloading}: Based on the estimated fit probabilities, the user chooses the best AI model with the highest fit probability and offload its tasks to the server possessing the AI model. 
\end{enumerate}

\subsubsection{\textbf{Model Training Process}}
On the server, model training process continues along with the task offloading process explained above. However, the training occurs in an offline manner, because the compact AI models are very sensitive to training. This means that the newly trained models are not used until their performance is verified. Eventually, the edge server trains and maintains its AI models to achieve higher overall fit probability with the users. These processes take more time; they take at most few hours to complete.  
\begin{enumerate} [a)]
\item \emph{Local model training}: The server uses the pilot data to train its local AI models to better suit the user’s needs. No additional training data is required, because they receive the pilot data, which is annotated and designed to represent an edge user’s tasks. Besides, the pilot data enables the edge server to reflect the temporal and spatial variations of the environment that the users will encounter, leading to increasing the corresponding fit probability. 
Depending on the complexity of the local models, the training may not be possible at a single edge server, which may have to be done collaboratively with other servers \cite{b2}, or offloaded to the cloud.  

\item \emph{AI model management}: A server always trains a copy of its AI model, because prematurely updating the trained version may lead to performance drop. When the older version of an AI model is clearly obsolete, the server replaces it with a newer version. Aside from managing the version of the models, the server also maintains different models with the same objective to maximize the AI model diversity gain. If necessary, a full universal model can be downloaded from the cloud to the edge server. 
\end{enumerate}

\section{Design of Energy Efficient Communication for Fit Probability Learning}\label{Sec: design}

To gain sufficient level of confidence in a fit probability, it is required for an edge user to upload a large amount of pilot data to the edge while consuming its limited energy.  It is thus vital to learn fit probability in an energy efficient manner, which is the main theme of this section. Although the design of energy efficient edge computing has been widely explored in the literature, our main focus is fundamentally different. Firstly, in the conventional MEC problems, the computation result of the offloaded task is assumed to be always exact. In contrary, the AI models in MoDNet introduce randomness to the results, which necessitates fit probability learning. Secondly, the number of tasks that need to be offloaded has been considered as a given value. Consequently, the problems are mostly concerned with deciding the proportions of locally computed and offloaded tasks. In contrast, the amount of pilot data is a control variable for problems regarding MoDNet, because it is directly related to the level of confidence in an AI model. These reasons call for designing new communication designs, which align well with the new trend of learning driven communication \cite{b12}. 

\subsection{Learning-Energy Tradeoff}

For learning fit probability, an edge user transmits a number of pilot data to the edge, which is used to sample the AI model's response as explained in Sec. \ref{Subsection:Architecture}. This acquisition of knowledge is mathematically expressed as the reduction of confidence intervals, which is the range of values that the actual fit probability lies with high likelihood. With more knowledge on the AI models, it becomes more likely to select the best AI model correctly.  However, it is detrimental when too much resource is used for learning the fit probabilities of AI models that are worse than the best AI model. This dilemma is commonly referred to as exploration-exploitation tradeoff, which is highlighted by \emph{multi-armed bandits} (MAB) problems~\cite{b13}.  

It is noteworthy that learning to explore and exploit well requires wireless transmission. According to Chernoff bound, the number of pilot data transmissions needed proportionally scales with the required confidence $\epsilon$, at the speed of $O(\log(1/\epsilon))$ \cite{b13}. Furthermore, the energy consumption super-linearly increases as the amount of transmitted data increases \cite{b14}. Combing the two makes the learning-energy tradeoff that we are exploring. 

\begin{figure}[t]
\centerline{\includegraphics[width=8cm]{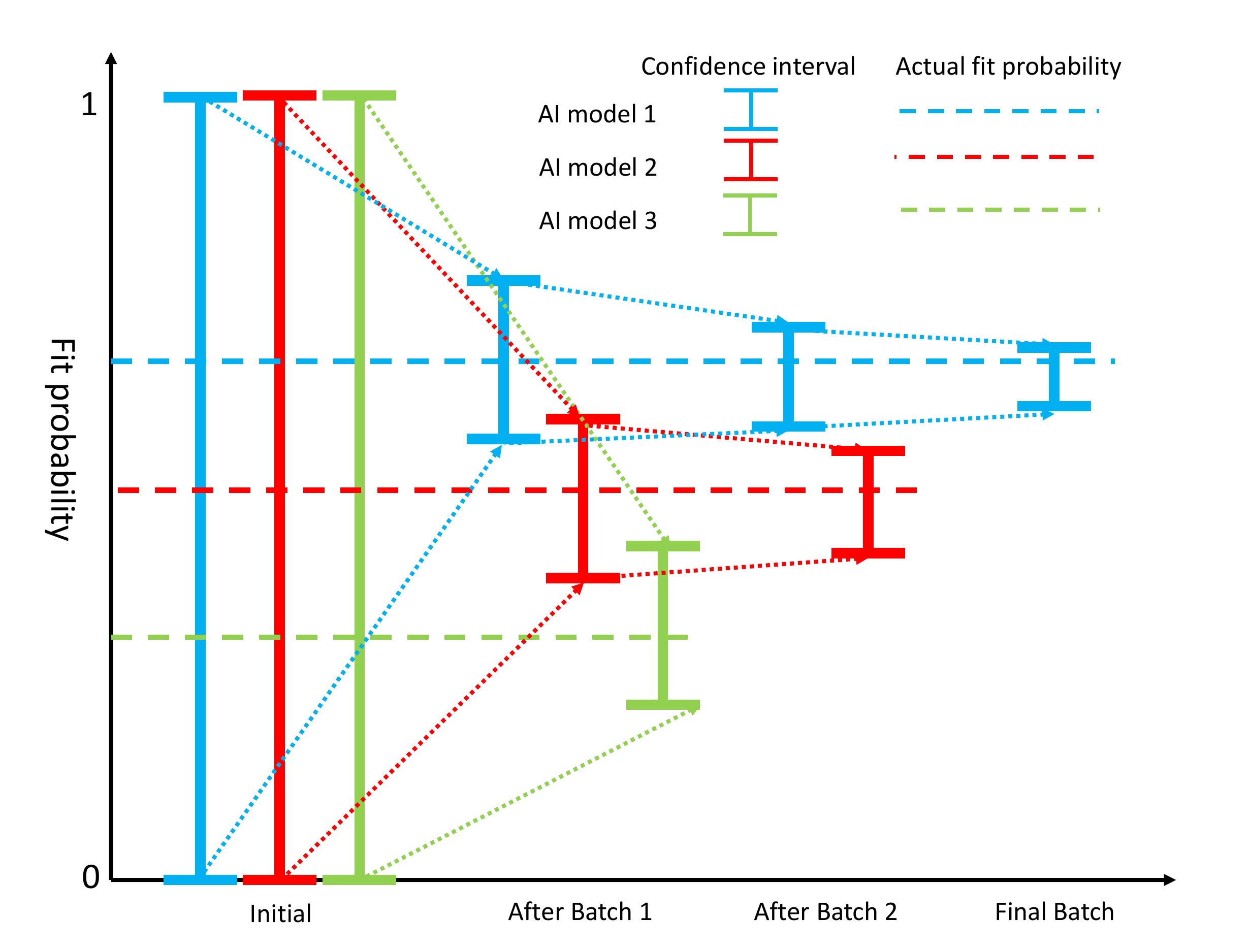}}
\caption{Graphical example of batch learning and selective elimination for fit probability learning.}
\label{fig5}
\end{figure}


\subsection{Batch Learning and Selective Elimination}

There are two ways for achieving higher energy efficiency. One is to increase the available transmission time to decrease the required data rate. The other is to decrease the total amount of data that have to be transmitted. The former is available by decreasing the signaling overheads, such as the feedbacks from the server containing the sampling results of the concerned AI model. Note that frequent feedbacks from the server induces more delay. Reducing the iterations of fit probability update leads to increasing the available transmission time. For these reasons, learning policies that transmit pilot data in batches are preferred \cite{b13}. However, if the batch size is too large, energy is wasted for learning the fit probabilities for all AI models, but the highest one is only required for the user. It is in conflict with the latter. 
Thus, dividing the batches into appropriate partitions is recommended. As illustrated in Fig.~\ref{fig5}, energy can be saved by observing intermediate learning results between batches and sequentially eliminating the AI models that perform seemingly poorly, i.e. its upper confidence bound is lower than the current best model’s lower confidence bound. In consequence, energy efficiency is determined by how accurately and quickly the inferior AI models are ruled~out. 

\subsection{New Research Directions }

Following the design guidelines mentioned above leads to revisiting several communication criteria regarded as de facto standards, to provide new research directions. 
\begin{itemize}
\item \textbf{Transmission mode selection}: Consider an extended edge network where multiple edge servers exist in the coverage of an edge user. To find a better AI model, the edge user can simultaneously broadcast the pilot data to multiple servers. However, it may result in excessive energy consumption, since the transmission rate is determined by the server whose channel gain is the worst. Things become worse if the AI models therein are unlikely to be selected due to their unfitness to user demand.  
As a result, the transmission mode should be adaptively controlled depending on the intermediate results of batched fit probability learning.  
\item \textbf{Opportunistic resource allocation}: In the case with multiple users, only a limited number of users are allowed to transmit the pilot data due to the shared radio resource. The opportunistic resource allocation that grants more resources to a user with a better channel gain may not help find the user with the best fit probability if the corresponding channel condition is bad. It is thus required to jointly consider channel gains and learning results to avoid wasting transmission energy.  
\item \textbf{User cooperation}: As the number of AI models or the number of edge servers increases, it could be a heavy burden for a user to learn the fit probabilities of all possible AI models. This burden is relieved by cooperating with several users whose AI model preference is similar. Specifically, by observing another user with similar model preference, a user can predict the fit probabilities of AI models that have not been learned yet. Measuring user similarity has been well studied in the recommendation system (see, e.g., \cite{b15}). More interestingly, a new tradeoff exists between users’ cooperation and similarity measurement thereof: for the users to cooperate efficiently, it is beneficial to make each user learn different AI models, but their similarity measurement becomes less accurate due to the decrease of overlapping knowledge.
\end{itemize}

\section{Concluding Remarks}
Uncertainty may be deemed as a dire issue for the edge. Nevertheless, devising progressive architectures grounded on the article’s principle and approach can turn the issue into a promising opportunity for the area of wireless communications. 
We start with guidelines for energy-efficient communication designs, but there are other promising research directions related to the efficient usage of different communication and computation resources. Seizing this opportunity depends on the novel efforts that are yet to come.       

\bibliographystyle{ieeetr}

\section*{Biography}

\noindent \textbf{Sejin Seo} is with the School of \emph{Electrical and Electronic Engineering} (EEE), Yonsei University. His current research interests include mobile edge computing, AI-assisted wireless communication, and related implementations. 

\noindent \textbf{Sang Won Choi} is  a senior researcher with the Korea Railroad Research Institute (KRRI). His research interests include mission-critical communications, mobile communications, communication signal processing, and information theory. 

\noindent \textbf{Sujin Kook} is  with the School of EEE, Yonsei University. Her current research interests include local 5G networks and AI-assisted wireless communication. 

\noindent \textbf{Seong-Lyun Kim} is  a Professor of wireless networks with the School of EEE, Yonsei University, Seoul, and the Head of the Robotic and Mobile Networks Laboratory and the Center for Flexible Radio. His research interests include radio resource management, information theory in wireless networks, collective intelligence, and robotic networks. 

\noindent \textbf{Seung-Woo Ko} (M’17) is  an assistant professor with Korea Maritime and Ocean University. His research interests include intelligent communications and computing, and localization. 

\end{document}